# Enabling older citizens' safe mobility: the ACTIVAGE approach

**Charis Chalkiadakis[1], Dimitris Tzanis[1], Evangelos Mitsakis[1]**

[1]Centre for Research and Technology Hellas – Hellenic Institute of Transport (CERTH-HIT)
E-mail: charcal@certh.gr

**Abstract**

We live in an ever-aging world. The percentage of older citizens increases in modern societies as older citizens represent the 19.20% of the general population. In Greece, an increase of almost 7% of older citizens has been observed in the last twenty years. As old age never comes alone, age-related impairments should be considered in the effort to provide safe transport conditions for them. It is of importance that transportation services have to meet the special requirements and needs of older citizens. European Union, through the ACTIVAGE project, aims at using the Internet of Things (IoT) solutions in favor of older citizens. In the framework of this project, the ACTIVAGE Safe Mobility Platform (ASMP) has been designed. Therefore, older citizens and their relatives have access to information regarding their travels. In the context of this study, an extensive description of the ASMP and the services offered through it is provided.

## *1. Introduction*

We live in an ever-aging world. The percentage of older citizens increases in modern societies with the median percentage of older citizens to be 19.20% of the general population. That percentage is the result of the increase of the elderly by almost 6% in the last twenty years. In Greece, an increase of almost 7% in the percentage of older citizens is observed for this period of time.

Because old age never comes alone, age-related impairments should be considered in the effort to provide safe transport conditions for older citizens. It is of importance that transportation services have to meet the special requirements and needs of older citizens. For that reason, significant impacts and challenges on the planning and provision of transport services arise. In order for the safe mobility of older citizens to be ensured, pre-trip and on-trip information has to be adapted to the needs of older citizens. European Union, through the ACTIVAGE project, aims at using the Internet of Things (IoT) solutions in favor of older citizens. Among others, a service related to the mobility of older citizens is included in the scope of the project.

In the framework of this project, the ACTIVAGE Safe Mobility Platform has been designed. The ACTIVAGE Safe Mobility Platform uses the network of Bluetooth detectors installed in the city of Thessaloniki to keep track of the mobility habits of the registered in the platform older citizens. Therefore, older citizens and their relatives have access to information like the mean speed, the meantime and the route of the said older citizens. There is also the provision of pre-trip information through the ACTIVAGE Safe Mobility Platform; routing services by taking into account historic data and information for the real-time congestion in the urban transport network are also provided to the registered users.

In the context of this study, an initial literature review regarding the age-related impairments of older citizens is provided. Afterward, there is an extensive description of the ACTIVAGE Safe Mobility Platform and the services offered through it.

## *2. Background*

The present chapter aims at providing an initial proof that older drivers are at high risk to involve in an accident; the risks of the said older drivers to have serious injuries in accidents is also a significant aspect mentioned in the literature.

Such issues are based in the age-related impairments older drivers may have and the percentages of accidents involving older drivers are expected to be higher in the next years due to the increase in the absolute number of older-aged people in the European Union (EU) and -even- worldwide.

### *2.1 Literature review*

According to studies found in the literature, older drivers have a higher risk of fatal and casualty crash involvement per distance travelled, relative to other drivers' age groups (Marottoli et al., 2005), (IET & ITS (UK), 2015), (Chandraratna & Stamatiadis, 2014), (OECD, 2001).

Older-aged drivers accounted for 25% of all traffic fatalities in 2013 in the EU (Polders et al., 2015). If older-aged persons are involved in a collision and their age is over 80, the possibilities of death (or of a serious injury) are double in relation to a younger driver (SaMERU, 2013). When the per distance crash rates were further investigated, it was found that there are two main factors leading to such results. Those factors are older drivers' physical frailty and hence vulnerability to injury in the event of a crash, and the fact that older drivers drive only short distances. The latter indicates a correlation between high crash involvement and low driving distance due to the fact that older drivers are not "in shape" in terms of executing the primary task of driving (Eby et al., 1998), (Marottoli et al., 2005), (IET & ITS (UK), 2015).

As proposed by the Institute of Engineering and Technology and the Intelligent Transport Systems (ITS) Association of the United Kingdom (UK) (IET & ITS(UK), 2015), the most effective way to avoid older drivers' injuries is to prevent the crash itself in the first place. One possible solution that may assist older drivers in executing with safety the primary task of driving is the use of in-vehicle ITS options. This is a potential solution to the older drivers' safety problem; however it is necessary to look at specific aspects of the crashes related to older drivers, in order to determine if these in-vehicle interventions or equipment options could be used efficiently to reduce the crash risk of older drivers (Whelan et al., 2006), (Polders et al., 2015).

The in-vehicle equipment capable of assisting the older drivers regarding driving (Najm et al., 1995), (Ling Suen et al., 1998), (Whelan et al., 2006) are presented in the following table (Table 1).

***Table 1:*** *Problems occurred by the age of older drivers and the assistance needed*

| Age-related impairments | Driving problems | In-vehicle equipment needed |
|---|---|---|
| Increased reaction time, difficulty dividing attention between tasks | Difficulty driving in unfamiliar or congested areas | Navigation/route guidance |
| Deteriorating vision, particularly at night | Difficulty seeing pedestrians and other objects – reading signs | Night vision/in-vehicle signs |
| Difficulty judging speed and distance | Failure to perceive conflicting vehicles and crashes at junctions | Collision warning |
| Difficulty perceiving and analyzing situations | Failure to comply with yield signs, traffic signals, and rail crossings, slow to appreciate hazard, difficulty in complex traffic maneuvers, such as lane changing and merging | In-vehicle signs and warnings, intelligent cruise control, automated lane changing and merging |
| Difficulty turning head/neck, reduced peripheral vision | Failure to notice obstacles | Blind spot detection, collision warning |
| More prone to fatigue | Get tired on long journeys | Intelligent cruise control, tired indicator |
| General effects of aging | Concerns over the inability to cope with a breakdown, driving to unfamiliar places, at night, in heavy traffic | Emergency callout (mayday) |
| Some impairments vary in severity from day to day | Concern over fitness to drive | Driver condition monitoring |

Concluding, it is observed in the literature that various guidelines have established, both in the United States of America (Stutts, 2005), (Potts et al., 2004) and EU (Polders et al., 2015), in order to facilitate the mobility conditions of older-aged drivers.

### *2.2 Statistics*

During the last years, a significant increase in the percentage of older citizens has been observed. In 2015, among the 7.3 billion people worldwide the older citizens correspond to 8.5% (or 617.1 million). In 2015, the median age in Europe was 41 years old and in 2050 it is expected to be 52 years old (Eurostat, 2015). Moreover, in 2016, older citizens (people of over 65 years old) constitute the 19.2% of the total population of the EU (Eurostat, 2015).

Figure 1 below depicts the percentage of older citizens, at EU level, as retrieved from Eurostat (Eurostat, 2015).

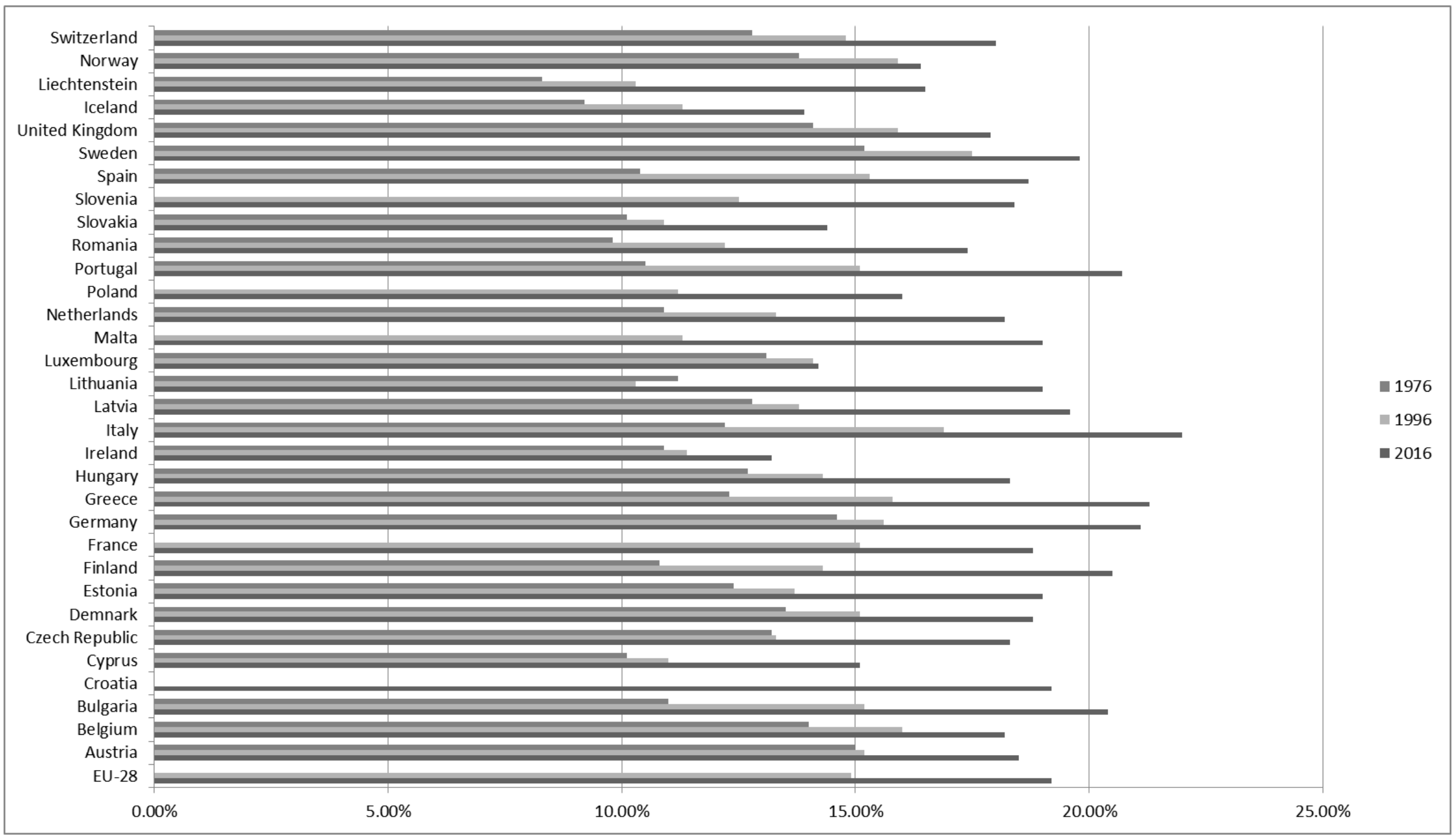

***Figure 1:*** *Elderly population as % of EU's total population*

Furthermore, statistics regarding the percentage of older citizens in the EU level for the years 2016-2018 are presented in the below figure (Figure 2). The information presented in Figure 2 is retrieved from Eurostat's database (Eurostat Statistics Explained).

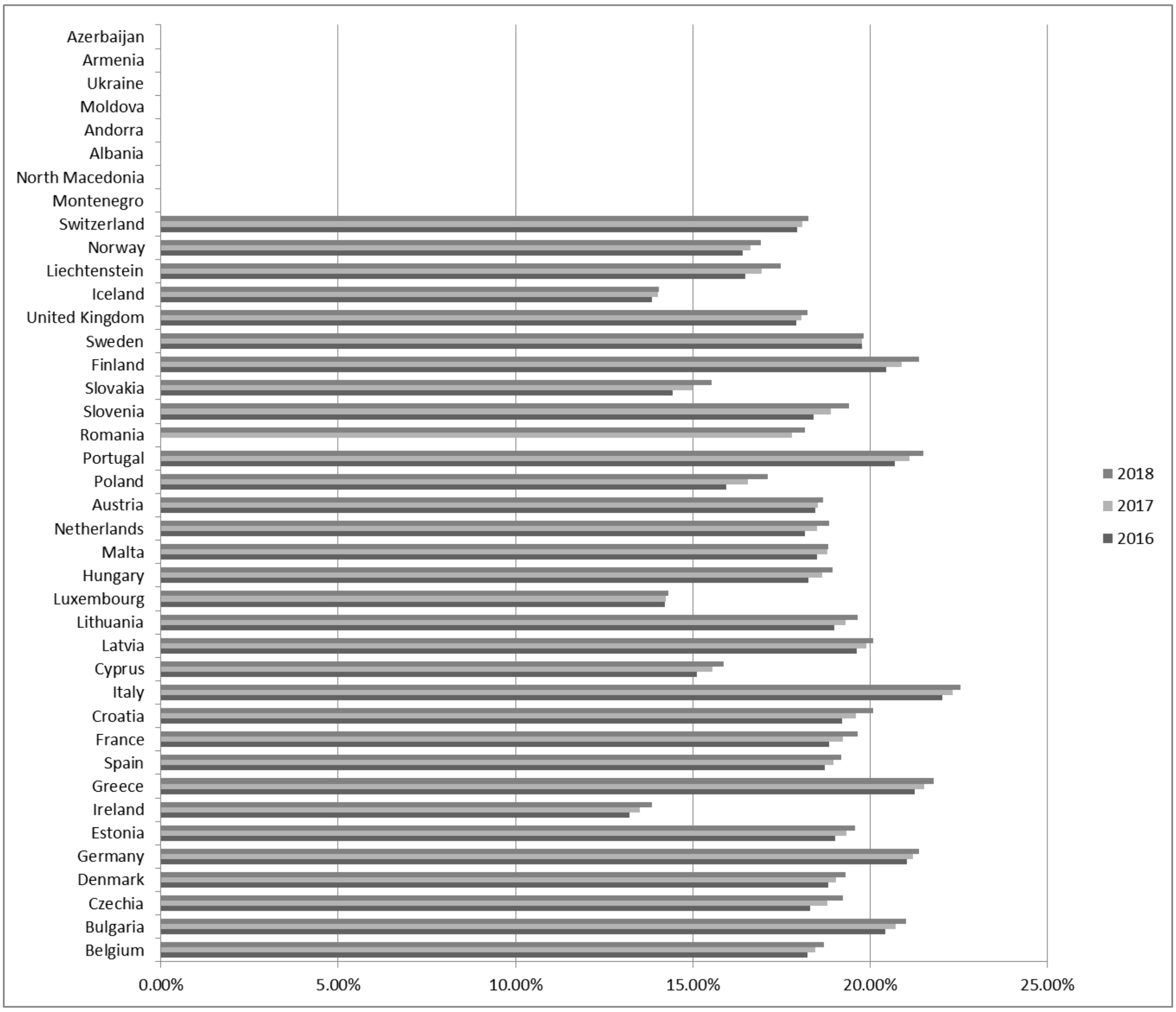

***Figure 2:** Elderly population as % of each country's total population*

In Figure 2, it is observed that Greece has the highest percentage of older citizens (in relation to Greece's population) among the other EU countries.

In the case of Greece, in 2016, older citizens constituted 21.4% of the total population and it is projected to increase more than 35% in the next 24 years. The expected increase of the elderly population (European Commission, 2017) is shown in the next figure (Figure 3).

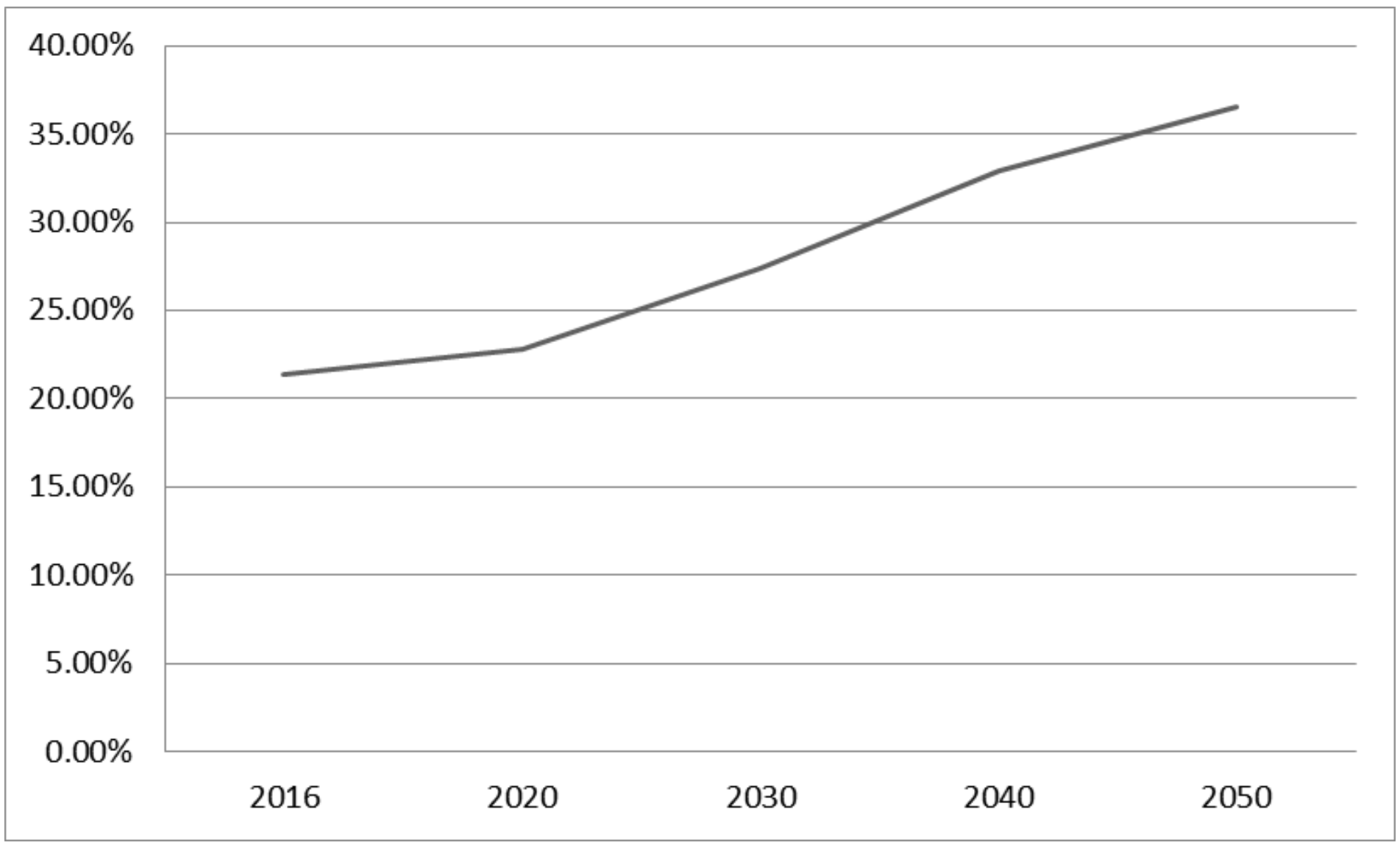

***Figure 3:*** *Elderly population as % of Greece's total population*

The projection presented above is confirmed from the statistics presented in Table 2 below (Eurostat, 2015), (European Commission, 2017).

***Table 2:*** *Older citizens' percentage in Greece and in the EU*

| | Greece value | EU mean value |
|---|---|---|
| 2016 | 21.26% | 18.16% |
| 2017 | 21.54% | 18.44% |
| 2018 | 21.79% | 18.74% |

According to the above, the overall increase in the number of older citizens both at EU and national level will lead to an ever-increasing need for transport systems which will have to keep pace with their needs. That will happen because, despite their advanced age, older citizens still have travel needs. Older citizens need to travel for social, cultural and physical activities as well as for accessing both neighborhood and non-neighborhood facilities and services. In addition, most of the problems faced by older people are interwoven with the ever reduced ability to travel from one place to another.

## *3. ACTIVAGE Safe Mobility Platform*

### *3.1 Introduction to the ACTIVAGE Safe Mobility Platform*

During the implementation of the ACTIVAGE project developed the ACTIVAGE Safe Mobility Platform. Through the ACTIVAGE Safe Mobility Platform three different use cases have been distinguished as the following Figure (Figure 4) shows.

| | Collect | Inform |
|---|---|---|
| Use Case 1: Mobility behavior monitoring | Frequency of travel (BT) | Frequency of travel, locations visited, modes of travel used, speed of travel |
| | Locations of travel (BT) | |
| | Mode of travel (FCD/GPS) | |
| | Speed of travel (BT & FCD/GPS) | |
| | Personal profile data | |
| Use Case 2: Pre-trip mobility information | Traffic conditions information (BT & FCD/GPS) | Comfort index information service |
| | Multimodal travel schedules | Multimodal routing service |
| | Parking availability information (Infotrip-Cosmos) | Multimodal information service |
| | Air quality information (Infotrip-Cosmos) | Parking availability service |
| | Personal profile data | |
| Use Case 3: Intersection alerts | Pedestrian presence monitoring at intersections (Infotrip-camera) | Alert driver when approaching an intersection of queue or red light |
| | Signal Phasing and Timing (SPAT) message | Alert driver when approaching an intersection of pedestrian crossing |

***Figure 4:*** *Use cases of the ACTIVAGE Safe Mobility Platform*

The main purpose of the creation and development of the ACTIVAGE Safe Mobility Platform is to enhance older drivers' safety, comfort, and awareness in the following three trip stages:

- Before trip.
- During trip.
- After trip.

According to the aforementioned trip stages, the following services are addressing the ageing population needs:

- *Pre-trip information* regarding the desired trip and its shortest path based on different parameters.
- *On-trip information* regarding intersection alerts.
- *Post-trip personalized information* and monitoring of the already executed trips.

In order, the above mentioned, purposes and needs of older drivers to be fulfilled/ accomplished, a specific platform (https://www.activage.imet.gr/) has been developed. Through the ACTIVAGE Safe Mobility Platform older citizens are provided with information regarding both their mobility habits and the traffic conditions of the urban and peri-urban transport network of Thessaloniki, thus covering pre-trip and post-trip information.

As the ACTIVAGE project aims at using the Internet of Things (IoT) solutions in favor of older citizens, the ACTIVAGE Safe Mobility Platform uses of the already installed Bluetooth Detectors (BT) (Mitsakis et al., 2015), (Mitsakis et al., 2017), in order to provide its registered users with a number of services. The ACTIVAGE Safe Mobility Platform provides its registered users (older citizens) with the following services:

- Trips Dashboard.
- Personal Trips.
- Traffic.
- Routing.

#### *3.1.1 Registration*

Older citizens, in order to have the ability to use the services provided through the ACTIVAGE Safe Mobility Platform, has to be registered.

During the registration process, the potential user states some personal information (Name, Surname. Father's Name, Date of Birth, Profession, Family Status, Contact Number, Address, Driving License, Car Owner), one or more Bluetooth MAC Addresses and the account details (Email and Password). The provided information and the collected data are encrypted, due to their sensitivity. That comes in terms with the latest General Data Protection Regulation (GDPR).

After the registration process, the user is able to have access to the ACTIVAGE Safe Mobility Platform's services.

#### *3.1.2 Trips Dashboard*

"Trip Dashboard" service aims at giving the users the opportunity of monitoring their trips either at a stated period of time. By using the stated (at the registration phase) Bluetooth MAC Addresses and the already installed BT, the users are provided with information regarding their mobility habits, at a stated period of time.

The provided information is illustrated in Figure (Figure 5) below.

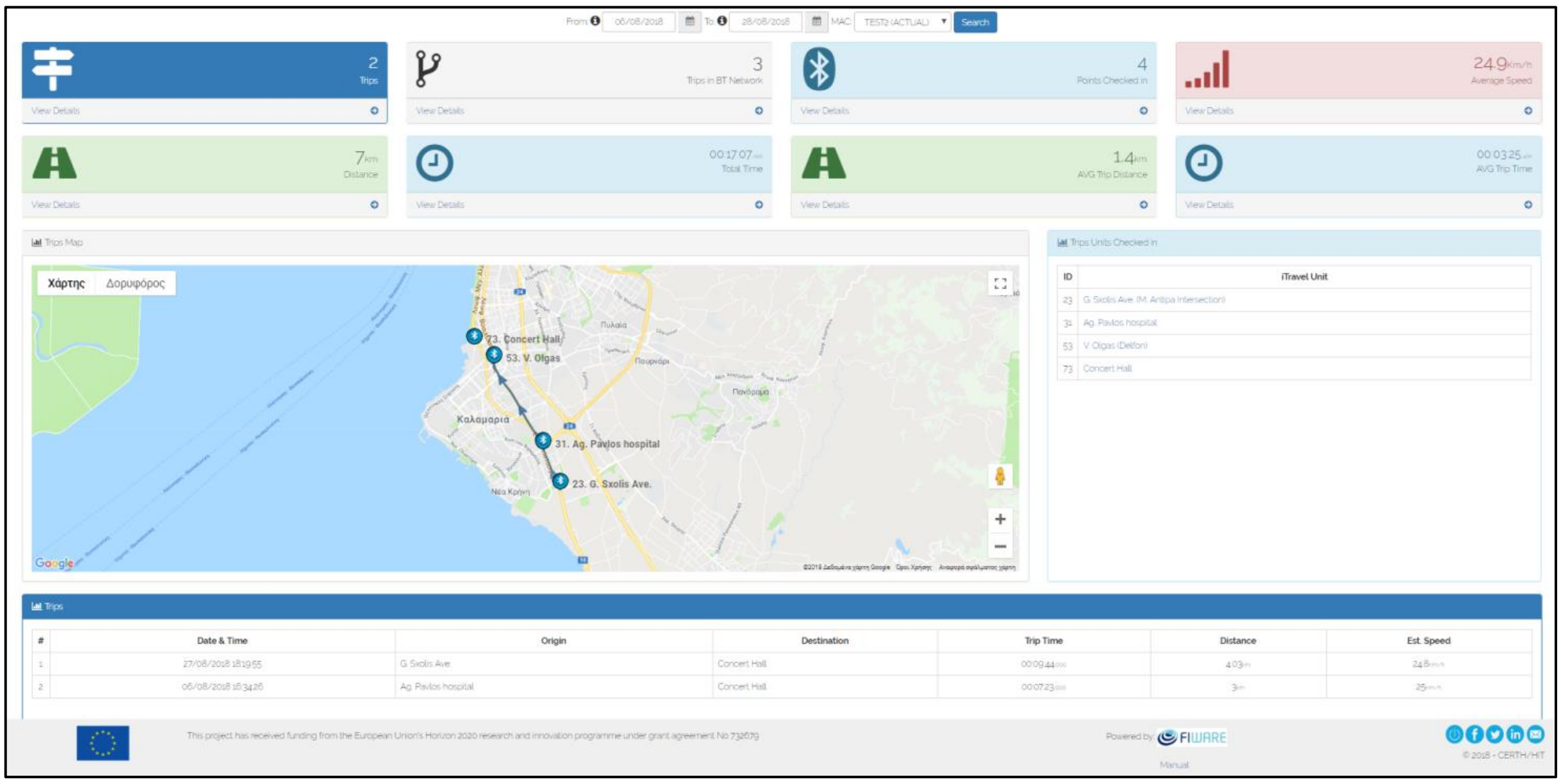

***Figure 5:*** *"Trips Dashboard" service interface*

For the stated time period (From 06/08/2018 – To 28/08/2018) and for the preferred MAC Address (TEST2 (ACTUAL)) information regarding the trips executed at this time period is provided. The user executed 2 trips (as presented under the "Trips" tab) while the user made three trips in the Bluetooth Detectors network and she/ he was checked-in in four points. The average speed was 24.9 kilometers per hour, the distance travelled was 7 kilometers, the total travel time was approximately 17 minutes and the average trip distance of the user (taking into account the total number of trips) is 1.4 kilometers.

### *3.1.3 Personal Trips*

"Personal Trips" service aims at giving the users the opportunity of monitoring their trips either at a stated period of time. By using the stated (at the registration phase) Bluetooth MAC Addresses and the already installed BT, the users are provided with information regarding their mobility habits, at a stated period of time. Moreover, "Personal Trips" service provides the users with additional detailed information regarding each trip separately.

For the same time period as above (From 06/08/2018 – To 28/08/2018), the "Personal Trips" service provides information regarding the Date & Time of the trip, the Origin, the Destination, the Trip Time, the Distance, the Estimated (Est.) Speed and the Comparison with actual average speed. There is also the possibility for the user to see specific information for each trip. This kind of information is similar to that provided through the "Trips Dashboard" service.

The provided information is illustrated in Figure (Figure 6) below.

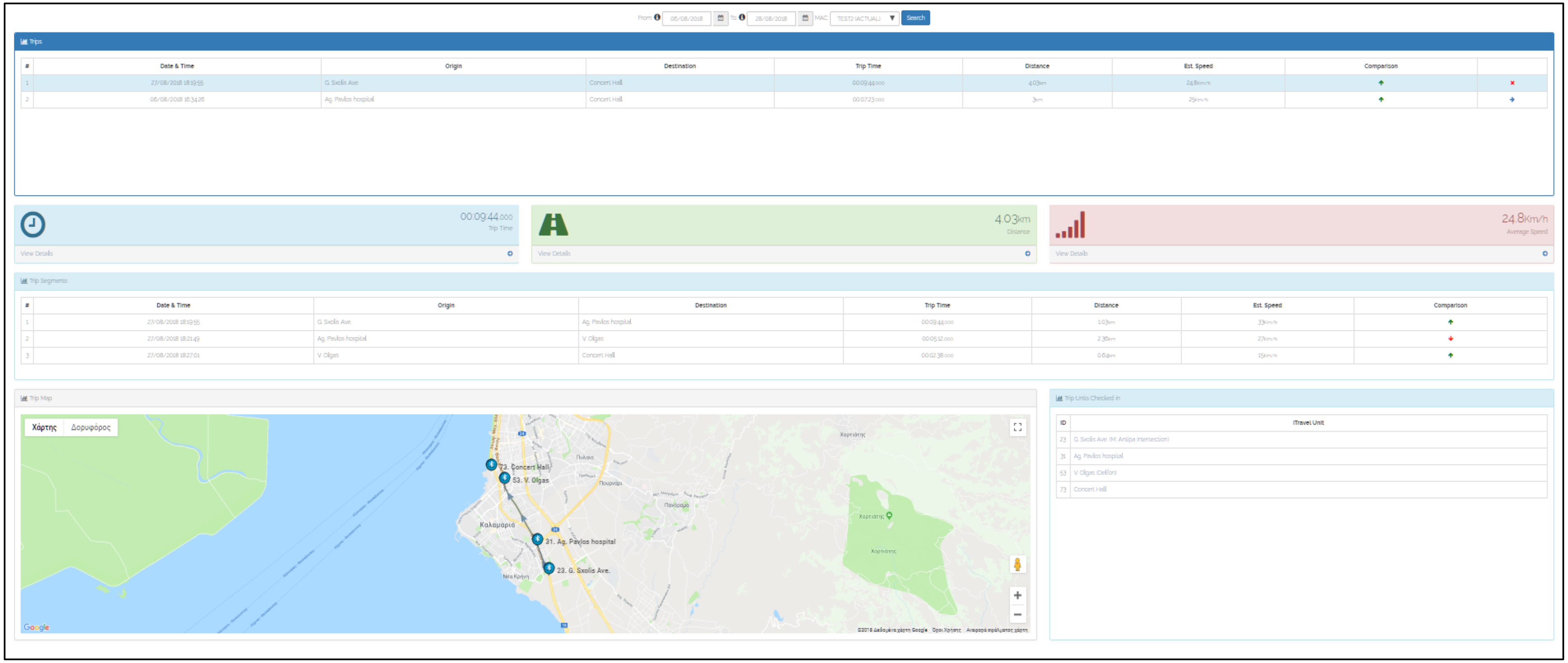

***Figure 6:*** *"Personal Trips" service interface*

### *3.1.4 Traffic*

The “Traffic” service provides information regarding the current traffic conditions in Thessaloniki’s urban and peri-urban transport network. The provision of this information is based on the collected data from the Bluetooth Detectors as well as on the Floating Car Data (FCD) that the Hellenic Institute of Transport (HIT) collects from taxi fleet. There is also the ability to provide information regarding the travel time and the traffic conditions in a number of pre-set point-to-point trips.

The provided information is illustrated in Figure (Figure 7) below.

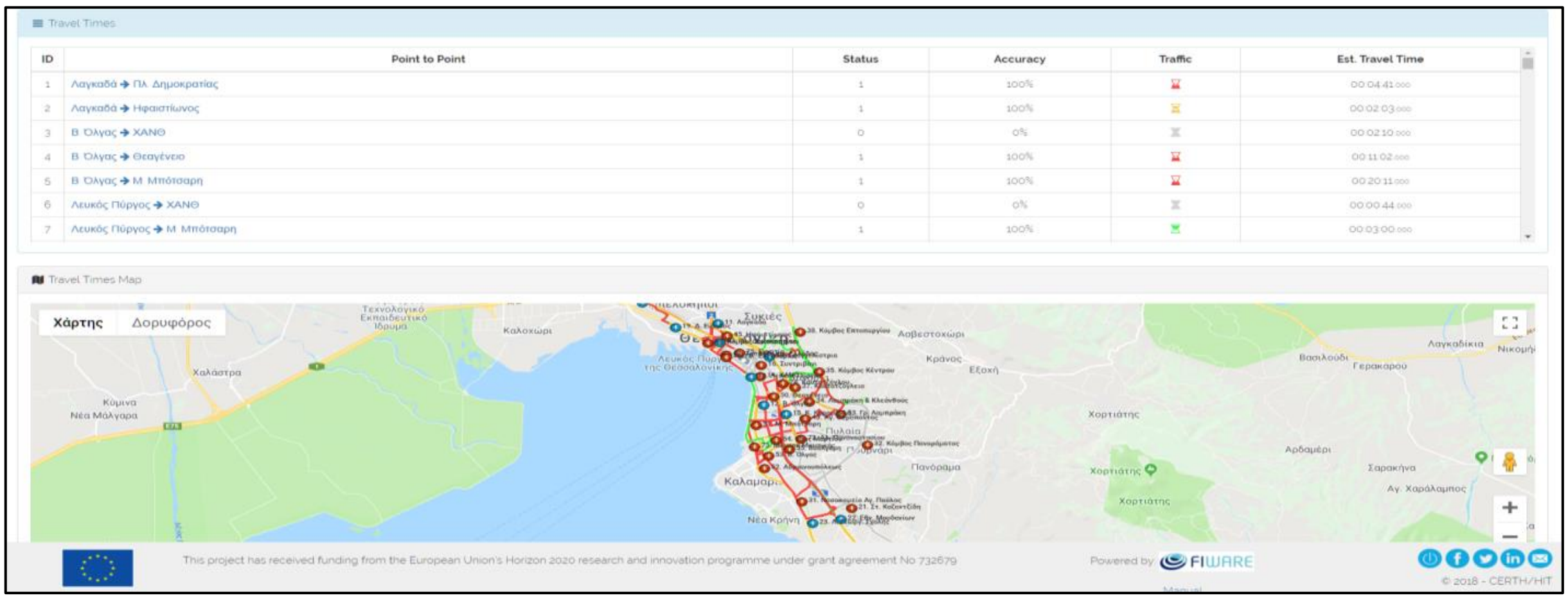

***Figure 7:** “Traffic” service interface*

### *3.1.5 Routing*

The “Routing” service allows the users to set their Origin and their Destination in order for the optimal route to be provided to them. There is the ability to choose whether the routing will be based either on real-time data (provision of the fastest route taking into account the current traffic conditions in the network) or on data that produce results with suggested routes with least travel time variability, thus increased comfort.

The provided information is illustrated in Figure (Figure 8) below.

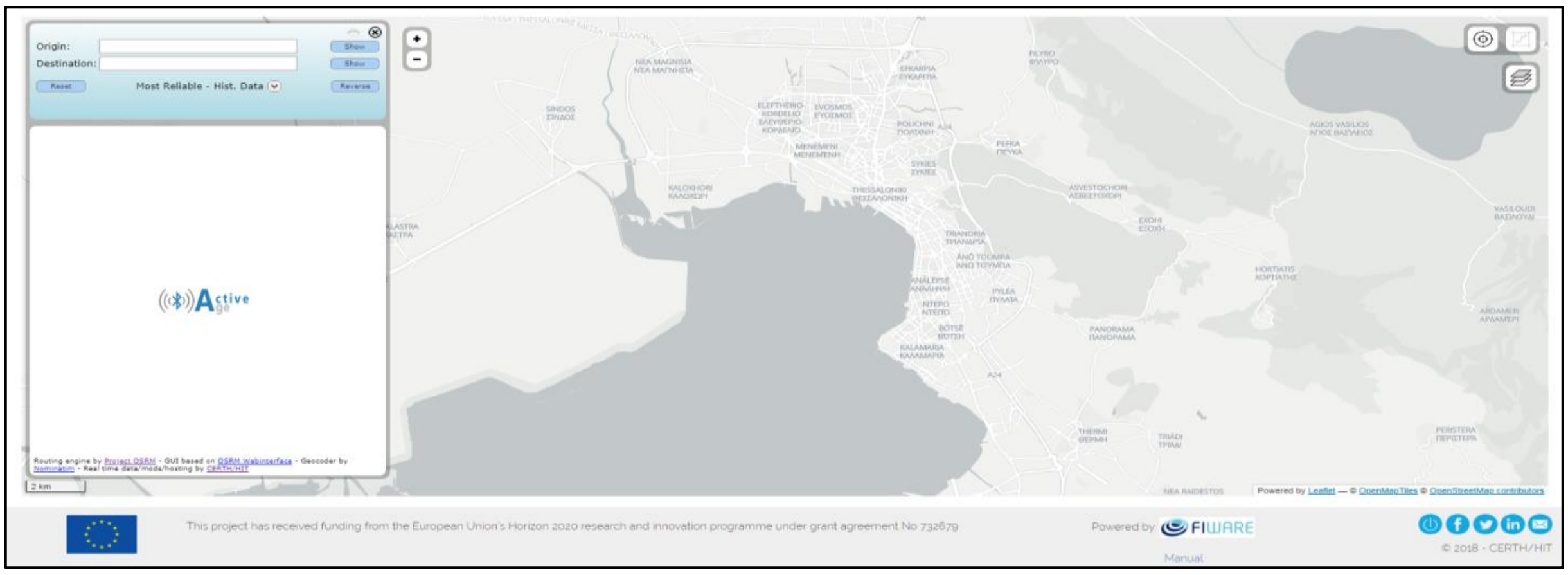

***Figure 8:** “Routing” service interface*

### *3.2 Specifications and Architecture of the ACTIVAGE Safe Mobility Platform*

ACTIVAGE Safe Mobility Platform has been designed and developed by the Hellenic Institute of Transport in the framework of the ACTIVAGE project. ACTIVAGE Safe Mobility Platform provides information regarding the following services:

1. Trip Dashboard
2. Personal Trips
3. Traffic
4. Routing

The aforementioned provided services are in line with the set Use Cases of the current study.

Two both discrete and collaborative data collection systems are used, in order for the proper collection and analysis of data. Data used is:

1. Personal data
2. Bluetooth detectors data

The two collected datasets are combined, in order for personalized data to be provided on the registered users of the ACTIVAGE Safe Mobility Platform. The following figure (Figure 9) presents in detail the overall architecture of the data collection and data analysis of the ACTIVAGE Safe Mobility Platform.

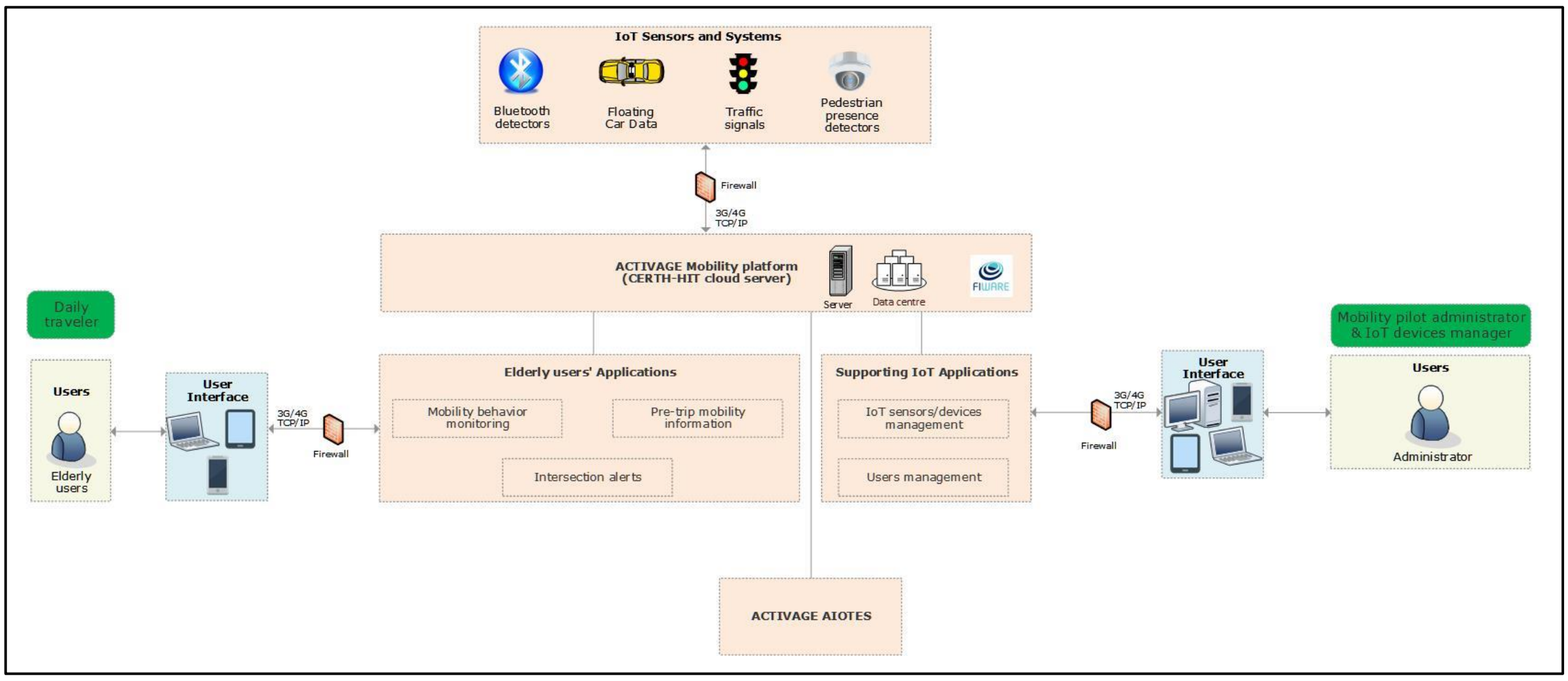

***Figure 9:*** *Architecture of the ACTIVAGE Safe Mobility Platform*

## *4. Validation Concept*

The use of the provided services aims to improve the mobility conditions of the older citizens in terms of comfort, safety, and awareness towards sustainable urban mobility. These factors are set as Key Performance Indicators which will be examined after a three-phase survey conducted among the users of the ACTIVAGE Safe Mobility Platform (Baseline survey, Intermediate survey, Post-Pilot survey).

It is expected, after the implementation of the services, to observe a better quality in the mobility conditions of the older citizens; it will be translated in an increase of the factors/ KPIs associated with comfort, safety, and awareness towards sustainable urban mobility. More specific:

***Table 3:*** *Key Performance Indicators (KPIs) and their associated target values*

| Factor | KPI | Target (increase in %) |
|---|---|---|
| Increased comfort of daily travel of older citizens through the provision of IoT-enabled infomobility services | Comfort | 10 |
| Increased travel safety of older citizens | Safety | 5 |
| Increased awareness towards sustainable urban mobility of older citizens | Awareness | 100 |

The pilot site is the wider metropolitan area of Thessaloniki. The wider urban and peri-urban road network of Thessaloniki is equipped with 40 existing Bluetooth detectors, as illustrated in Figure 10.

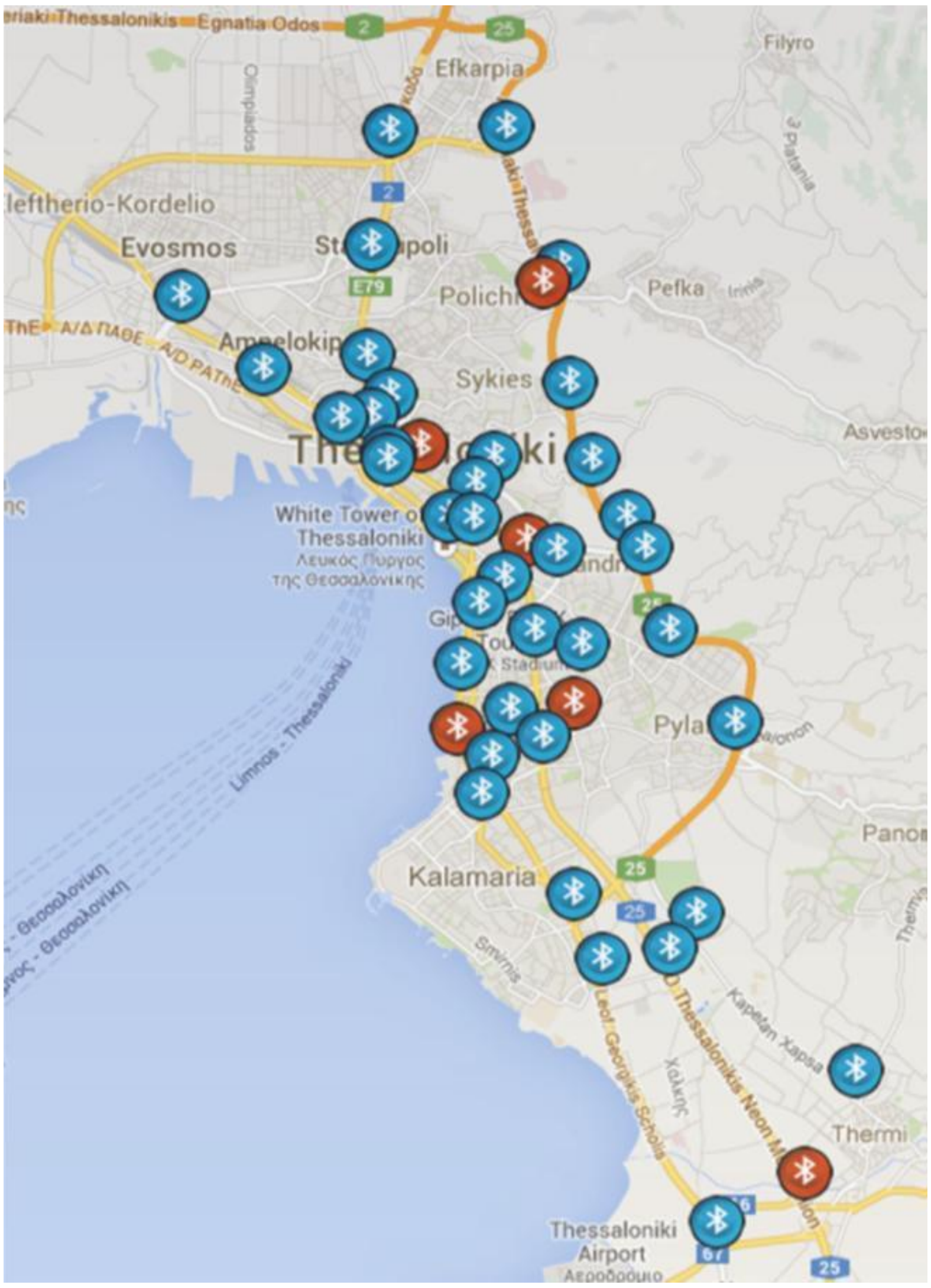

***Figure 10:*** *Existing Bluetooth detectors network*

The acceptability of the IoT based system will be extracted as an outcome of the conducted baseline, intermediate and post-pilot surveys.

As described above, in order for the KPIs to be measured a survey will be conducted. The survey consists of three discrete phases (Baseline, Intermediate, Post-Pilot) for both the better understanding of the usefulness of the platform and the evaluation of the services.

The Baseline survey consists of 16 questions, which aim to give feedback regarding both the perceived mobility conditions of the older citizens, based on the current situation (no use of the ACTIVAGE Safe Mobility Platform's services) and the expected usability of the provided services. The Intermediate and the Post-Pilot surveys both consist of 35 questions and they aim at providing feedback for the perceived mobility conditions and the usability of the platform during the use of the platform and after the end of the pilot phase. The outcome of the services' evaluation will provide information regarding the three KPIs.

## *5. Conclusions*

Older drivers face significant difficulties related to their mobility. The source of such difficulties is mainly the age of the driver and, therefore, the age-related impairments appeared in people of advanced age. The said impairments set the physical integrity and, in most of the cases, the life of older aged drivers in jeopardy. This is expected to increase in the following years, considering the ever-growing percentage of older aged persons in the EU.

Taking these facts into account, it is important to provide all the necessary equipment to this group of vulnerable users, in order to improve their mobility conditions. In this context, the ACTIVAGE project aims at using IoT technologies in favor of older citizens, in general, and of older drivers particularly.

In the framework of the ACTIVAGE project, a platform (ACTIVAGE Safe Mobility Platform) has been designed in order to provide the registered in the platform older-aged drivers with information concerning both their mobility habits, as well as real-time traffic conditions of the urban and peri-urban transport network of Thessaloniki.

As a way forward, the collection of data from the registered users of the ACTIVAGE Safe Mobility Platform will provide further insight in the acceptability and the usefulness of the provided information both with the quantification of the set KPIs and further statistical analysis of the questionnaires filled from the platform users. Finally, the collected data are expected to provide all the necessary information for further analysis of the mobility patterns of older drivers.in the city of Thessaloniki.

## *6. References-Bibliography*